\documentclass[11pt,leqno]{article}   % 11pt,

\usepackage[T1]{fontenc}
\usepackage[utf8]{inputenc}
\usepackage{authblk}
\usepackage[all]{xy}
\usepackage{float}
\usepackage{amsmath,amstext,amsthm,amsfonts,amssymb,color,latexsym}
\numberwithin{equation}{section}
\usepackage{epsfig}
\usepackage{subfig}
\usepackage{caption}

\title{Transition Energy in Strong and Weak Confinement of Type-I Spherical Core-shell Quantum Dots}
\author[a]{T. Shelawati,\footnote{shelatiansin@gmail.com}} 
\author[a,b]{M. S. Nurisya\footnote{risya@upm.edu.my}}
\author[a,b]{C. Kar Tim}
\author[a]{A. K. Mazliana}
\affil[a]{Department of Physics, Faculty of Science, Universiti Putra Malaysia, 43400 UPM Serdang, Selangor, Malaysia}
\affil[b]{Laboratory of Computational Sciences \& Mathematical Physics, Institute for Mathematical Research, Universiti Putra Malaysia, 43400 UPM Serdang, Selangor, Malaysia}
\usepackage{hyperref}
\usepackage{cleveref}
\begin{document}

\maketitle
\begin{abstract}
In this paper, we present a mathematical approach in estimating transition energy of  type-I core-shell quantum dots (CSQDs) in strong and weak confinement of charge carriers within the core materials. We will  do this by using effective-mass approximation together with a single-band model. The effects of potential step of conduction band and valence band on confinement strength will be discussed. It can be seen at the end of this paper that for a same size CSQDs, CSQDs with bigger potential steps will have stronger carriers confinement and with more localized excitons. \\

\end{abstract}

\section{Introduction}

In recent years, research field has witnessed vigorous study in nanoscale semiconductors, ranging from quantum wells (2-dimensional), quantum wires (1-dimensional) and quantum dots (0-dimensional), where their charge carriers are spatially confined in one degree, two degrees and three degrees respectively. Quantum dots, or sometimes dubbed as "artificial atoms" or "hyperatoms"~\cite{A32} due to their size reaching atomic scale, leading to discrete energy levels like atoms~\cite{QDH}, have shown great potential in energy, agriculture and medicine fields as sensors~\cite{A01}, solar panels \cite{A01,A05,A23}, bio-tagging materials \cite{A04,A05,A08,A23} or light-emitting devices \cite{A05,A23}. It has long established that their nano-scale size that is comparable to Bohr radius gives rise to a strong localization of charge carriers compared to their bulk counterparts. The small size property of quantum dots causes it to have a high surface to volume ratio (quantum dots with radius of 5 nm may still have $10^4$ atoms), made its surface to play a significant role, which sometimes can be considered as a drawback due to possibility of having defects on its surface \cite{A08, A10}. But many research have discovered that it is possible to cap or encapsulate these quantum dots surfaces with another organic or inorganic semiconductor materials in order to enhance their optical and electrical properties as described in \cite{A23,A26,A29,A33}. This novel finding, known as core-shell quantum dots (CSQDs) was further studied to distinguish their different types based on region of confined electrons and holes and their band edges alignment. For type-I CSQDs, the charge carriers are confined only in core material or only in shell material (inverse type-I CSQDs). As for type-II CSQDs, the charge carriers are confined between conduction band of core material and valence band of shell material or between conduction band of shell material and valence band of core material (inverse type-II). There are three well-known ways of heteroepitaxial growth for CSQDs fabrication which are adding shell compound layer-by-layer method onto the core surface (Frank-van der Merwe), layer-by-layer followed with additional islands formation (Stranski-Krastanov) or just islands formation directly on the core surface (Volmer-Weber). All these passivating modes are governed by their interface energies and lattice mismatch \cite{A34, QDH}.

The motivations of passivating quantum dots surfaces are to provide electrical and chemical passivation \cite{A04,A08} that will improve the photostability of the quantum dots against oxidation \cite{A04,A10} as well as to increase the confinement of charge carriers and making its transition energy more tunable \cite{A01,A04,A06,A08,A10}.  Inorganic capping materials can also reduce the chemical activity due to cationic and anionic dangling bonds \cite{A08}. But increasing the thickness of this shell layer above its critical thickness will cause the shell to impair the whole core-shell quantum dots instead of improving its photoluminescene \cite{A34}. Hence, a study to investigate relation between shell thickness and transition energy of CSQDs is very crucial. Different approaches had been introduced such as single-band model \cite{A01, A02, A04, A06, A08, A10, A23}, multiband theory (8-band model) \cite{A02, A22, A11}, envelope function \cite{A04}, density-functional theory (DFT) \cite{A05} and model-solid theory \cite{A01}. In this paper, on top of applying the single-band model onto our CSQDs, we will also introduce two different types of confinement; strong confinement and weak confinement model that are based on strengths of potential steps on conduction and valence bands, instead of just depending on their core sizes as suggested in \cite{A32} and \cite{QDH}.

This paper is organized where we will first discuss the transition energy in bare quantum dots, and followed with explanation on single-band approach on energy transition in CSQDs. Then, we will apply this mathematical approach on CSQDs with strong confinement, followed with CSQDs with weak confinement. Subsequently, we will discuss the result by observing the radial probabilities in both strong and weak confinement cases.

\section{Theory}

\subsection{Transition energy for core Quantum Dots}

\begin{figure}[H]
	\center 														   	\includegraphics[width=5cm]{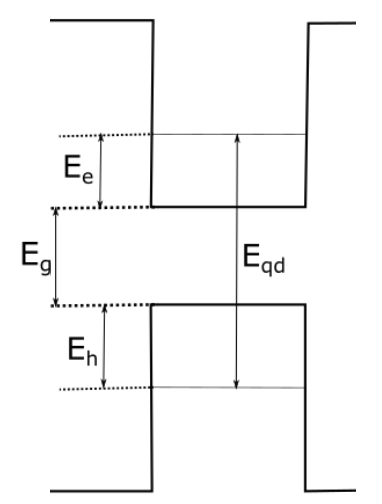} 	
	\caption{Schematic diagram of transition energy in quantum dots.}   
	\label{fig:transition}							
\end{figure}

Following the effective mass approximation (EMA), both electrons and holes of quantum dots will have their specific effective mass that is direction-dependent (anisotropic), rather than having free electron mass. In Figure \ref{fig:transition}, we can see that total transition energy for core QDs, $E_{qd}$ is the total of exciton energy ($E_h + E_e$) and the bulk band gap energy ($E_g$).

The effective Hamiltonian for exciton will then have the form as in Equation \ref{eqd}.

\begin{equation}
\label{eqd}
H_{\text{exc}}=-\frac{h}{2m_e}\bigtriangledown_{e}^{2}-\frac{h}{2m_h}\bigtriangledown_{h}^{2}-\frac{e^2}{4 \pi \varepsilon_o \varepsilon_r |r_e-r_h|} + \text{polarization terms} .
\end{equation}

The solution of eigenvalue for the above Hamiltonian can be written down as

\begin{equation}
\label{eqd2}
E_\text{exc}=\frac{\pi^2 \hbar^2}{2r^2}(\frac{1}{m_e}+\frac{1}{m_h})-\frac{1.8 e^2}{4 \pi \varepsilon_o \varepsilon_r r} + \text{small terms},
\end{equation}

where $m_e$ is effective mass for electrons, $m_h$ is effective mass for holes, $r$ is the radius of quantum dots, $\varepsilon_o$ is vacuum permittivity and $\varepsilon_r$ is dielectric constant of our material. Note that, for bare quantum dots, it has $1/r^2$ dependence which implies that as the core quantum dots getting smaller, the lowest transition energy will increase due to increase in confinement energy that rises due to strong localization of electrons and holes. Here, we treated the Coulomb interaction term ($\frac{e^2}{4 \pi \varepsilon |r_e+r_h|}$) as first order perturbation that rises due to interactions between charge carriers. Our lowest eigenvalue is not very sensitive with this term (and often neglected) as it has $1/r$ dependence against our confinement energy term that has $1/r^2$ \cite{A01,A28}. But for more general cases and realistic model, we will include this term as integral over spherically-symmetry surface of the quantum dots. 

%------------------------------------------------
\subsection{Single-band model}

We consider our colloidal CSQDs as spherical core quantum dots encapsulated by a layer of a different compounds of semiconductor. As we increase the shell thickness, we are actually imposing a strain on the core quantum dots with step potential. We first have to separate our wavefunction into radial and angular part as in Equation \ref{eqn:separate}.

\begin{equation}
\label{eqn:separate}
\Psi_{nlm}(\vec{r})=R_{nl}(r)Y_{lm}(\theta,\phi),
\end{equation}

where $n$, $l$ and $m$ are principal, orbital and magnetic quantum numbers respectively.

We are only interested in solving the radial wavefunction for lowest transitions energy ($n=1$ and $l=m=0$), which is transition between lowest unoccupied molecular orbital (LUMO) of conduction band and highest occupied molecular orbital (HOMO) of valence band, in other words $1s_e\rightarrow 1s_h $ transition. To solve this, we could benefit from spherical Bessel functions (for case $E_{e,h} > V_{c,v}$) or modified spherical Bessel funtions (for case  $E_{e,h} < V_{c,v}$) \cite{A01}.

\begin{figure}[H]
\center
\label{fig:bands}
\includegraphics[scale=0.7]{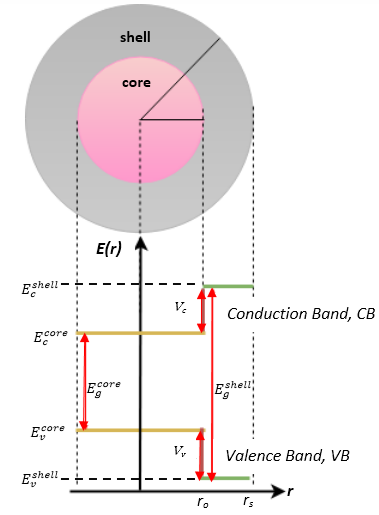}
\caption{Schematic diagram of band line-up of core-shell QDs.}
\end{figure}

$V_c$ and $V_v$ are potential steps for conduction band and valence band respectively, $r_o$ and $r_s$ are radius of core quantum dots and radius of CSQDs repectively. Our potential energies $V(r)$ are step-like potentials and going to infinity outside shell layer. 

\begin{align*}
V(r)=\left\{\begin{matrix}
0   &; &0\leqslant r\leq r_o\\ 
V_{c,v} &; &r_o\leqslant r\leq r_s\\
\infty &; &r\geq r_s .
\end{matrix}\right.
\end{align*}

For $E_{e,h} > V_{c,v}$

\begin{align}
\label{eqn:weakcon}
R_i(r)=\left\{\begin{matrix}
R_1(r) =A j_l(k_1r)   &; & 0\leqslant r\leqslant r_o\\ 
R_2(r) =B j_l(k_2r)+C n_l(k_2r)) &; & r_o\leqslant r\leqslant r_s\\
R_3(r) =0 &; & r\geqslant r_s.
\end{matrix}\right.
\end{align}

For $E_{e,h} < V_{c,v}$ 

\begin{align}
\label{eqn:strongcon}
R_i(r)=\left\{\begin{matrix}
R_1(r) =A j_l(k_1r)   &; & 0\leqslant r\leqslant r_o\\ 
R_2(r) =B i_l(k_2r)+C \kappa_l(k_2r)) &; & r_o\leqslant r\leqslant r_s\\
R_3(r) =0 &; & r\geqslant r_s.
\end{matrix}\right.
\end{align}

Note that $j_l(k_i)$, $n_l(k_i)$, $i_l(k_i)$ and $\kappa_l(k_i)$ are spherical Bessel function of first kind, Neumann spherical functions, modified  spherical Bessel function of first kind and modified spherical Bessel function of second kind respectively.  $k_1=\frac{\sqrt{2m^*_1E_{e,h}}}{\hbar}$ and $k_2=\frac{\sqrt{2m^*_2[V_{c,v}(r)-E_{e,h}]}}{\hbar}$ for $E_{e,h} < V_{c,v}$ or $k_2=\frac{\sqrt{2m^*_2[E_{e,h} - V_{c,v}(r)]}}{\hbar}$ for $E_{e,h} > V_{c,v}$ and  $m_i^*$ is effective mass of electrons, $m^e_i$ or effective mass of holes, $m^h_i$. Our boundary conditions require us to have 

\begin{align*}
R_i (r) &= R_{i+1}(r) \\
\dfrac{1}{m_i}R'_i (r) &= \dfrac{1}{m_{i+1}}R'_{i+1}(r),
\end{align*}

where $i=1$ for core, $i=2$ for shell and $i=3$ for outside shell layer. 

Following the said boundary conditions, Equation \ref{eqn:weakcon} and Equation  \ref{eqn:strongcon} will form three equations that subsequently will solve for normalization constants \textit{A}, \textit{B} and \textit{C} that have to satisfy $\int_{0}^{r} 4 \pi r^{2}R_l^{2}(r)dr=1$. \\

\begin{equation}
\label{eqn:det1}
\begin{bmatrix}
j_0(k_1r_o) & -j_0(k_2r_o) & -n_0(k_2r_o)\\ 
\frac{1}{m_1^*}j_0'(k_1r_o) & -\frac{1}{m_2^*}j_0'(k_2r_o) & -\frac{1}{m_2^*}n_0'(k_2r_o)\\ 
0 & j_0(k_2r_s) & n_0(k_2r_s)
\end{bmatrix} 
\begin{bmatrix}
A\\ 
B\\ 
C
\end{bmatrix}= 0 
\end{equation}

\begin{equation}
\label{eqn:det2}
\begin{bmatrix}
j_0(k_1r_o) & -i_0(k_2r_o) & -\kappa_0(k_2r_o)\\ 
\frac{1}{m_1^*}j_0'(k_1r_o) & -\frac{1}{m_2^*}i_0'(k_2r_o) & -\frac{1}{m_2^*}\kappa_0'(k_2r_o)\\ 
0 & i_0(k_2r_s) & \kappa_0(k_2r_s)
\end{bmatrix} 
\begin{bmatrix}
A\\ 
B\\ 
C
\end{bmatrix}= 0 
\end{equation}

Equation \ref{eqn:det1} and \ref{eqn:det2} will form transcendental equations that obey $ det(E_{e,h})=0$. Our transcendental  equations then can be written as 

\begin{equation}
\label{eqn:trans1}
\dfrac{m_2^*}{m_1^*}\dfrac{j'_0(k_1r_o)}{j_0(k_1r_o)}=\dfrac{j'_0(k_2r_o)n_0(k_2r_s)- j_0(k_2r_s) n'_0(k_2r_o)}{j_0(k_2r_o)n_0(k_2r_s)- j_0(k_2r_s) n_0(k_2r_o)}, 
\end{equation}

or

\begin{equation}
\label{eqn:trans2}
\dfrac{m_2^*}{m_1^*}\dfrac{j'_0(k_1r_o)}{j_0(k_1r_o)}=\dfrac{i'_0(k_2r_o)\kappa_0(k_2r_s)- i_0(k_2r_s) \kappa'_0(k_2r_o)}{i_0(k_2r_o)\kappa_0(k_2r_s)- i_0(k_2r_s) \kappa_0(k_2r_o)}.
\end{equation}

One needs to solve Equation \ref{eqn:trans1} or \ref{eqn:trans2} in order to find $E_{e,h}$. As we need to solve for both electron energy and hole energy separately, hence the use of superscript will be handy to remind us whether we are working on the electron states $(R_i^e)$ or hole states $(R_i^h)$. 

%---------------------------

\section{Application in CSQDs with Strong Confinement}

For this part, we will be applying the above approach to a CSQDs model that has a narrow core that encapsulated with a much wider band gap shell compound, where the step-potentials will be bigger than the core bulk band gap ($V_{c,v} \gg E_g^{bulk}$),  to imply a strong excitons confinement within the core material. For this purpose, we chose PbS/CdS core-shell QDs with material parameters as shown in Table \ref{table:2}. The transition energy for CSQDs will be

\begin{equation}
E_\text{qd}=E_g^{bulk} + E_e + E_h -\frac{1.8 e^2}{4 \pi \varepsilon_o \varepsilon_r r_o}. 
\end{equation}

\begin{table}[H]
\caption{List of parameters for PbS/CdS core-shell QDs.}
\centering
 \begin{tabular} {|c |c |c |} 
 \hline 
  Materials & PbS \cite{Z02} & CdS \cite{Z02} \cite{B12} \\
 \hline
 Band gap (eV) & 0.41 &   2.5 \\ [9pt]\hline
 $m_e/m_0$ & 0.08 &  0.2 \\ [9pt] \hline
$m_h/m_0$ & 0.08 &  0.7  \\ [9pt] \hline
Dielectric constant, $\varepsilon_r$ & 169 &  8.73  \\ [9pt] \hline
$V_c$ (eV) & \multicolumn{2}{|c|}{1.254}  \\ [9pt] \hline
$V_v$ (eV) & \multicolumn{2}{|c|}{0.836}   \\ [9pt] \hline 
 \end{tabular}
\label{table:2}
\end{table}

\begin{figure}[H]
\centering
\begin{minipage}{.55\textwidth}
  \centering
  \includegraphics[width=1\linewidth]{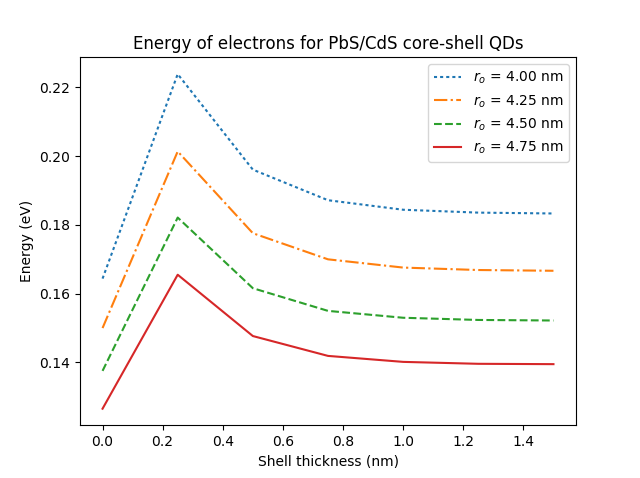}
  \captionof{figure}{Energy of electrons for
  PbS/CdS.}
  \label{fig:pbscds1}
\end{minipage}%
\begin{minipage}{.55\textwidth}
  \centering
  \includegraphics[width=1\linewidth]{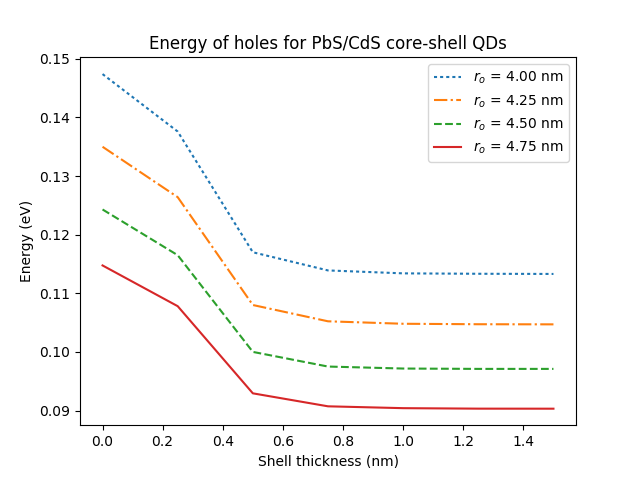}
  \captionof{figure}{Energy of holes for PbS/CdS.}
  \label{fig:pbscds2}
\end{minipage}
\end{figure}

\begin{figure}[H]
\centering
\includegraphics[scale=0.6]{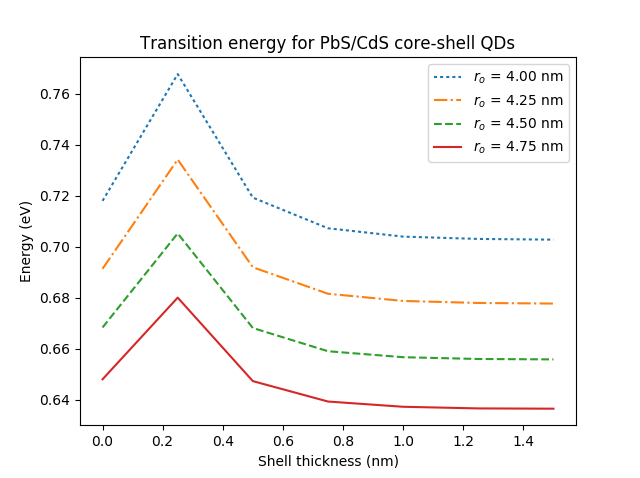}
\caption{Transition energy for PbS/CdS CSQDs.}
\label{fig:pbscds3}
\end{figure}

\section{Application in CSQDs with Weak Confinement}

For this second part, we will be applying the single-band model to a wide core QDs encapsulated with a slightly wider shell compound, where the step-potentials will be smaller than the core bulk band gap ($V_{c,v} \ll E_g^{bulk}$) to imply a weak excitons confinement within the core material. For this purpose, we choose ZnTe/ZnSe core-shell QDs with material parameters as shown in Table \ref{table:3}.

\begin{table}[H]
\caption{List of parameters for ZnTe/ZnSe core-shell QDs.}
\centering
 \begin{tabular} {|c |c |c |} 
 \hline
 Materials & ZnTe \cite{Z02} \cite{A23} & ZnSe\cite{Z02} \cite{A23} \\
 \hline
 Band gap (eV) & 2.394 &   2.8215 \\ [9pt]\hline
 $m_e/m_0$ & 0.11 &  0.14 \\ [9pt] \hline
$m_h/m_0$ & 0.7 &  0.6  \\ [9pt] \hline
Dielectric constant, $\varepsilon_r$& 8.7 &  9.1  \\ [9pt] \hline
$V_c$ (eV) & \multicolumn{2}{|c|}{0.2565}  \\ [9pt] \hline
$V_v$ (eV) & \multicolumn{2}{|c|}{0.171}   \\ [9pt] \hline
 \end{tabular}
\label{table:3}
\end{table}

\begin{figure}[H]
\centering
\begin{minipage}{.55\textwidth}
  \centering
  \includegraphics[width=1\linewidth]{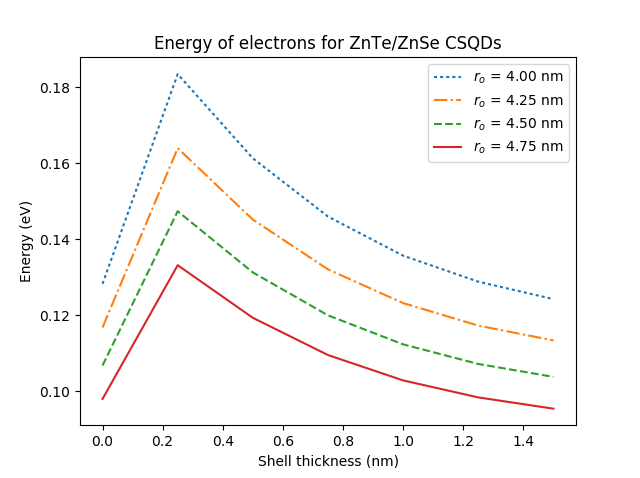}
  \captionof{figure}{Energy of electrons for
  ZnTe/ZnSe.}
  \label{fig:znteznse1}
\end{minipage}%
\begin{minipage}{.55\textwidth}
  \centering
  \includegraphics[width=1\linewidth]{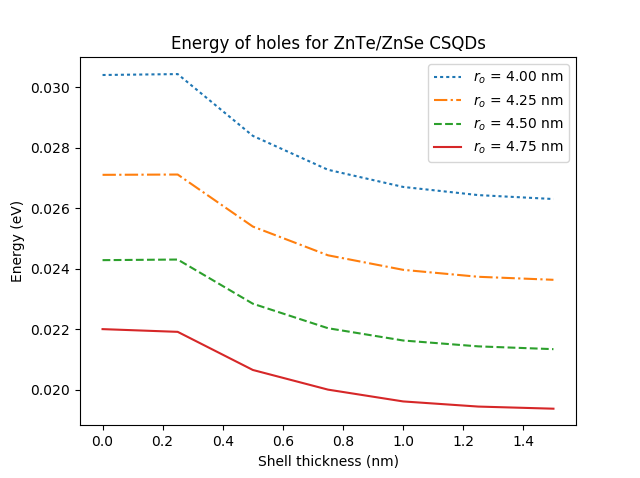}
  \captionof{figure}{Energy of holes for ZnTe/ZnSe.}
  \label{fig:znteznse2}
\end{minipage}
\end{figure}

\begin{figure}[H]
\centering
\includegraphics[scale=0.6]{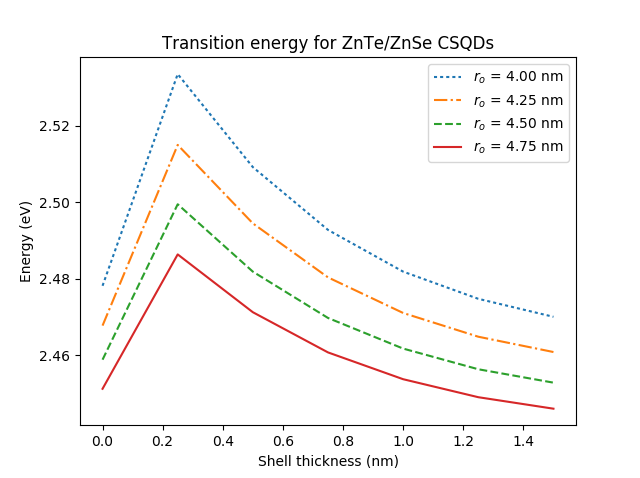}
\caption{Transition energy for ZnTe/ZnSe CSQDs.}
\label{fig:znteznse3}
\end{figure}

%------------------------------------------------

\section{Analysis on Radial Probability}

\subsection{Strong Confinement Case}

\begin{figure}[H]
\centering
\includegraphics[scale=0.7]{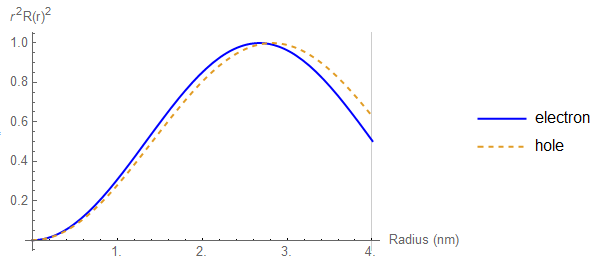}
\includegraphics[scale=0.7]{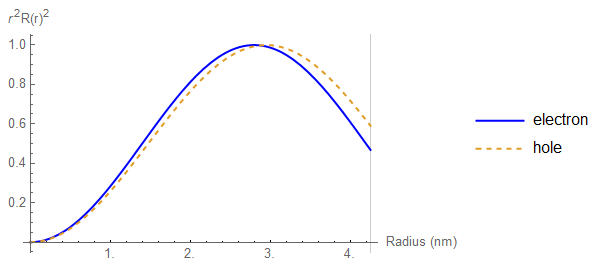}
\caption{Radial probability of carriers in (top) 4.00 nm PbS quantum dots and (bottom) 4.25 nm PbS quantum dots.}
\label{fig:corePbS}
\end{figure}

For case of bare PbS quantum dots with radius of 4.00 nm and 4.25 nm, there is no significant difference in radial probability of finding charge carriers within the cores as shown in Figure \ref{fig:corePbS}, although the transition energy is higher in 4.00 nm quantum dots compared to 4.25 nm quantum dots.

\begin{figure}[H]
\centering
\includegraphics[scale=0.7]{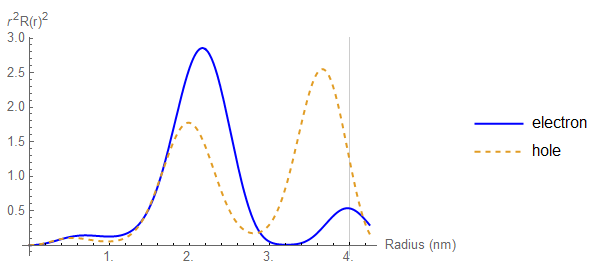}
\caption{Radial probability of carriers in 4.25 nm PbS/CdS with 4.00 nm core radius.}
\label{fig:csPbS1}
\end{figure}

After adding 0.25 nm of CdS shell layer onto the 4.00 nm PbS quantum dots (CSQDs radius = 4.25 nm), we can see in Figure \ref{fig:csPbS1}, that the radial probability of charge carriers are shifted more to the center of CSQDs, signaling a stronger localization compared to the bare 4.25 nm PbS quantum dots. This stronger carriers localization which translate into stronger carriers confinement is proved with higher transition energy in 4.25 nm PbS/CdS CSQDs (4.00 nm core radius with 0.25 nm shell thickness) compared to transition energy in 4.25 nm bare PbS quantum dots. 

\begin{figure}[H]
\centering
\includegraphics[scale=0.7]{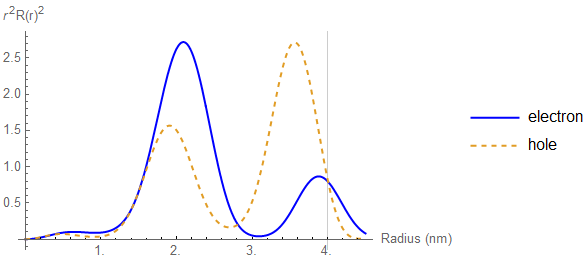}
\caption{Radial probability of carriers in 4.50 nm PbS/CdS with 4.00 nm core radius.}
\label{fig:csPbS2}
\end{figure}

But as we increase the thickness layer of CdS, the radial probability of finding the charge carriers started to spread into the shell region, as shown in Figure \ref{fig:csPbS2}. This spreading is caused by the defects in interface area, causing delocalization of charge carriers where it becomes easier for charge carrier to tunnel through the interface and be in shell region.

\subsection{Weak Confinement Case}

For bare ZnTe quantum dots that has wide bulk band gap, the confinement of charge carrier is not too obvious, leading to smaller increasing factor in transition energy compared to narrow band gap case. In narrow band gap, the energy for quantum dots almost doubled, while in wide band gap, the increasing is less pronounced. It is worthy to note that group II-VI semiconductor is less likely to have a strong carriers confinement \cite{QDH, A32}.

\begin{figure}[H]
\centering
\includegraphics[scale=0.7]{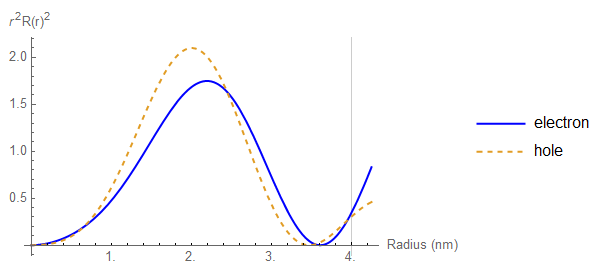}
\caption{Radial probability of carriers in 4.25 nm ZnTe/ZnSe with 4.00 nm core radius.}
\label{fig:csznte1}
\end{figure}

But after adding 0.25 nm of ZnSe shell layer, the transition energy of ZnTe/ZnSe CSQDs  is increasing by more than 5 $\%$ compare to its bare quantum dots. It also can be seen from Figure \ref{fig:csznte1} that the probabilities of the carriers in shell region are relatively higher  compared with strong confinement case as in Figure \ref{fig:csPbS1}. This shows weaker localization of charge carriers.

\begin{figure}[H]
\centering
\includegraphics[scale=0.7]{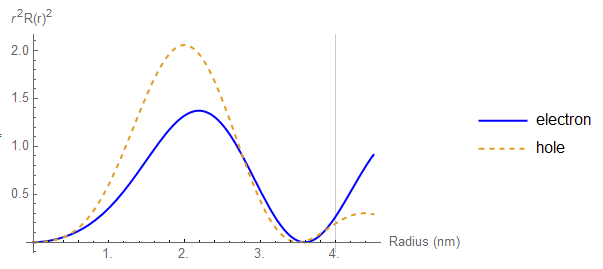}
\caption{Radial probability of carriers in 4.50 nm ZnTe/ZnSe with 4.00 nm core radius.}
\label{fig:csznte2}
\end{figure}

The probability of charge carrier is spreading more into shell region as shown in Figure \ref{fig:csznte2} as we increase thickness shell layer, due to strong strain energy imposed on the core quantum dots, causing lattice mismatch to be significant in order to minimise the strain on interface area \cite{A34}.

\begin{figure}[H]
\centering
\includegraphics[scale=0.7]{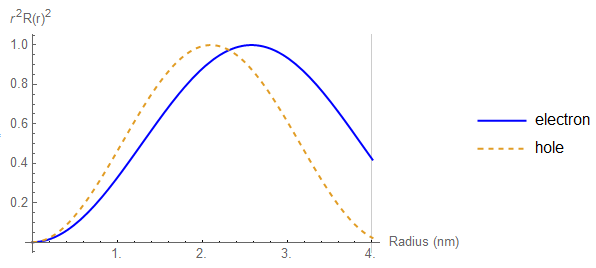}
\caption{Radial probability of carriers in 4.00 nm ZnTe quantum dots.}
\label{fig:coreznte}
\end{figure}

%------------------------------------------------

\section{Conclusion}

We can see that even with a same CSQDs size, we can still have different types of confinement; a strong confinement and a weak confinement of carriers as we impose a big and small potential step respectively. This is important factor to consider for fabrication of CSQDs as instead of relying on size factor to tune the confinement strength, we can also control it by manipulating the potential step. Increasing shell thickness will increase photoluminescene of CSQDs as we basically increase confinement of electrons and holes \cite{A08}. But as it reaching its critical thickness, the strain on interface will be significant enough to induce lattice mismatch in order to minimize the strain \cite{A34}. The high energy excitons now can tunnel easily through the core-shell interface, causing further delocalization, hence slowly reducing the confinement strength. As the delocalization of carriers keeps decreasing, they become less influenced by the shell thickness \cite{A08}, as shown in Figure \ref{fig:pbscds3} and Figure \ref{fig:znteznse3}.

%------------------------------------------------

\section{Acknowledgment}

We would like to thank Ministry of Education Malaysia for supporting this research through the Fundamental Research Grant Scheme (FRGS/1/2018/STG02/UPM/02/10).
%----------------------------------------------------------------------------------------
%	REFERENCE LIST
%----------------------------------------------------------------------------------------

%----------------------------------------------------------------------------------------

\end{document}